\documentclass[twocolumn]{aa}
\usepackage{graphicx}
\usepackage{txfonts}
\begin{document}
   \title{Detection of Giant Pulses in pulsar \object{PSR J1752+2359}}

   \author{A.~A. Ershov
          \inst{1}
          \inst{2}
          \and
          A.~D. Kuzmin
          \inst{1}
          \inst{2}}

   \offprints{A.~A. Ershov,
   \email{ershov@prao.psn.ru}}

   \institute{Pushchino Radio Astronomy Observatory, Astro Space
              Center, Lebedev Physical Institute,
              Russian Academy of Sciences, Pushchino, 142290,
              Russia
        \and Isaac Newton Institute, Chile, Pushchino Branch, Russia}

   \date{Received 29 November 2004 / Accepted 20 July 2005}

   \abstract{
   We report the detection of Giant Pulses (GPs) in the pulsar
\object{PSR J1752+2359}. About one pulse in 270 has a peak flux
density more than 40 times the peak flux density of an average
pulse (AP), and the strongest GP is as large as 260. The energy of
the strongest GP exceeds the energy of the average pulse by a
factor of 200 which is greater than in other known pulsars with GPs.
PSR J1752+2359 as well as the previously detected pulsars
PSR B0031--07 and PSR B1112+50, belong to the first group of
pulsars found to have GPs without a strong magnetic field at
the light cylinder.

    }

   \authorrunning{A.~A. Ershov and A.~D. Kuzmin}
   \titlerunning{Detection of Giant Pulses in pulsar PSR J1752+2359}

   \maketitle

\section{Introduction}

Giant pulses (GPs) are short duration burst-like increases of the
intensity of individual pulses from pulsars.

The peak intensities and energies of GPs greatly exceed the peak
intensity and energy of the average pulse (AP).  The energy
distribution  of GPs follows a power-law. The GPs are much
narrower than the APs and their phases stably placed within the
APs.

This rare phenomenon was first detected in the Crab pulsar
\object{PSR B0531+21} (Staelin \& Sutton \cite{staelin}) and the
millisecond pulsar \object{PSR B1937+21} (Wolszczan et al.
\cite{wolszczan}), both with very strong magnetic fields on the
light cylinder of $B_{\rm LC} = 10^4 - 10^5$~G. This gave rise to
the suggestion that GPs occur in pulsars with very strong magnetic
fields at the light cylinder and a search of GPs was oriented
towards those pulsars. As a result GPs were detected in five other
pulsars with very strong magnetic fields at the light cylinder :
\object{PSR B0218+42} (Joshi et al. \cite{joshi}), \object{PSR
B0540--69} (Johnston \& Romani \cite{johnston}), \object{PSR
B1821--24} (Romani \& Johnston \cite{romani}), \object{PSR
J1823--3021} (Knight et al. \cite{knight})  and \object{PSR
B1957+20} (Joshi et al. \cite{joshi}).

We report the detection of GPs in the pulsar PSR J1752+2359.
Correlating this with our previously published data on PSR
B1112+50 (Ershov \& Kuzmin \cite{ershov}) and PSR B0031-07 (Kuzmin
et al. \cite{kuzmin04}), we show that GPs exist in pulsars with
relatively low magnetic fields at the light cylinder of $B_{\rm
LC} = 1~-~100$~G.

The detection of GPs in PSR J1752+2359 is a result of our
monitoring program for GPs. More details will be published upon
completion of this program.

\section{Observations}

Observations were performed from December 24, 2003 to October 04,
2004 with the Large Phase Array (BSA) Radio Telescope at Pushchino
Radio Astronomy Observatory of the Lebedev Physical Institute at a
frequency of 111 MHz. BSA is a transit telescope with an effective
area of about 15\,000 square meters. One linear polarization was
received.  We used a 128-channel receiver with a channel bandwidth
of 20 kHz. The  sampling interval was 2.56 ms and the receiver
time constant was $\tau = 3$~ms.

All observations were time-referenced to the Observatory's
rubidium master clock, which in turn was monitored against the
National Time standard via TV timing signals. During the off-line
data reduction, the signal records were cleaned of radio
interferences. Subsequently, the inter-channel dispersion delays
imposed by the interstellar medium were removed. Each observation
was analyzed for pulses with amplitudes exceeding a preset level,
and its amplitude, pulse width and phase were derived.

For verification that GPs were actually detected we used similar
detection technique to  McLaughlin \& Cordes (\cite{mclaughlin})
and Cordes et al. (\cite{cordes}) for the detection of GPs. Our
processing consisted of: 1) checks for radio interference, 2)
de-dispersing by summing over frequency channels while taking into
account the time delay associated with plasma dispersion in the
interstellar matter, 3) averaging the time series synchronously
with the pulsar period to form a standard intensity profile, 4)
identifying individual giant pulses and their occurrence times by
selecting intensity samples that exceedreed the off-pulse mean by 5
sigma, 5) aligning average profiles and individual giant pulse
profiles in pulse phase with the
TimApr\footnote{psun32.prao.psn.ru/olegd/soft.html} program. For
the TimApr analysis, we used the timing model from Lewandowski et
al. (\cite{lewandowski}).

\section{Results}

\begin{figure}
     \centering
      \resizebox{\hsize}{!}{\includegraphics{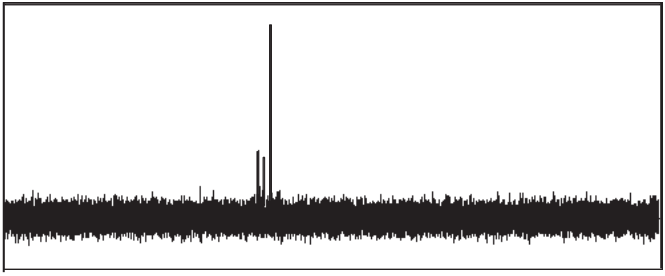}}
      \caption{
One observation session of GPs. Three large pulses stand out of
the noise  background and weak pulses are observed inside 420
pulsar periods.
              }
         \label{Fig1}
   \end{figure}

\begin{figure}
     \centering
      \resizebox{\hsize}{!}{\includegraphics{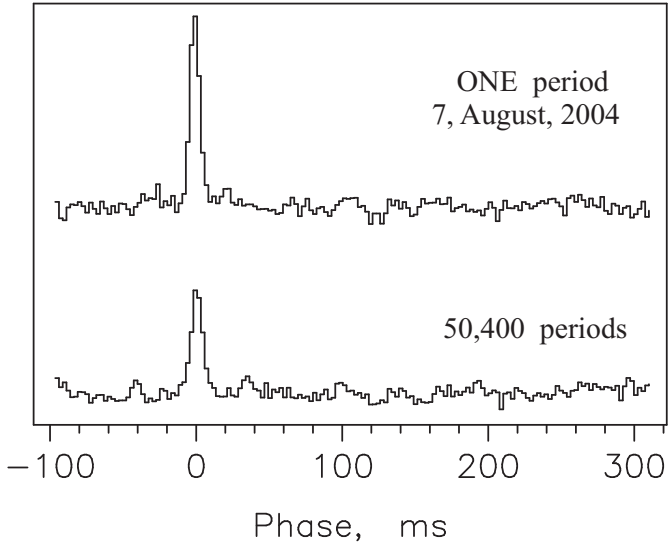}}
      \caption{
\textbf{(Top)} The strongest observed pulse. \textbf{(Bottom)} The
average pulse profile containing 50\,400 pulsar periods. The
intensity of the profiles is shown in arbitrary units.
              }
         \label{Fig2}
   \end{figure}

Only the brightest pulses are well separated from the underlying
pulse and noise fluctuations. They are both single pulses and
groups of several strong pulses. Fig.~\ref{Fig1}  shows an example
of one  observation session of GPs. A group of three bright pulses
standing out of the noise  background and underlying weak pulses
were observed inside 420 pulsar periods.
%
%
The 187 pulses (1 pulse for 270 observed periods) with $S/N \geq
5$ were selected and analyzed. The observed peak flux density
exceeded the peak flux density of the average pulse  (AP) (which
is equivalent to the integrated profile) by more than a factor of
40. Figure~\ref{Fig2} shows the strongest observed GP together
with the AP averaged over 50\,400 pulsar periods.
%
%

The value of the GP flux density was determined relative to the AP
flux density as
$$
  S_{\rm peak}^{\rm GP} = S_{\rm peak}^{\rm AP} \times (I^{\rm GP} /I^{\rm AP})~,
$$
where $I^{\rm GP} /I^{\rm AP}$ is the ratio of the GP to AP peak
intensity and $S_{\rm peak}^{\rm AP}$ is the peak flux density of
AP
$$
  S_{\rm peak}^{\rm AP} = S_{\rm mean}^{\rm AP} /k_{\rm form}~,
$$
where $S_{\rm mean}^{\rm AP}$ is the flux density averaged over a
pulsar period $P$, $k_{\rm form} = w_{\rm eff}/P $ is the duty cycle
of the pulsar, and $w_{\rm eff}$ - the effective pulse width.

The value of $S_{\rm mean}^{\rm AP}$ was measured relative to the
reference discrete source 3C452, and to the reference pulsars PSR
B0329+54, B0809+74, B0823+26, B0834+05, B0919+06, B1541+09,
B1839+56 and B1919+21 with known flux densities.

To normalize the observation system's parameters and control its
stability, we have performed control observations of the reference
pulsar PSR B1919+21 throughout the same time interval as the GPs.
Its flux density, measured in the same way as for PSR J1752+2359,
is stable within 25 percent. We calculate the AP of the total
intensity by summing six partial-sum profiles of 20 observation
sessions (8400 pulsar periods) in which signal-to-noise ratio was
larger than five. These samples revealed similar values of flux
density, within 15 percent. The AP mean flux density of PSR
J1752+2359 was $S_{1752} = 11~\pm~4$~mJy.

Interstellar scintillations (ISS) do not significantly affect our
intensity measurements. At the frequency of our observations (111
MHz), the time scales of the refractive ISS for both J1752+2359
and PSR B1919+21 ($DM = 36$ and $12.4 ~\mathrm{pc~cm^{-3}}$) are
about 1 min (Malofeev et al. \cite{malofeev}), and smoothed over
the duration of our observation of several minutes. The long term
refractive ISS are weak and do not seriously affect the intensity
measurements.

The peak flux density of the AP is $S_{\rm peak} = S_{1752} /
k_{\rm form}$, the pulsar period is $P = 409$~ms and the effective
width of the AP is $w_{\rm eff} = 11.3$~ms. Thus $k_{\rm form} =
0.0276$ and the peak flux density of the AP is $S_{\rm peak} =
400$~mJy.

The observed peak flux density of the strongest GP exceeds the
peak flux density of the AP by a factor of 260. Therefore its peak
flux density is  $S_{\rm peak}^{\rm GP} = 105$~Jy.

The energy of the  strongest observed GP is $S_{\rm peak}^{\rm GP}
\times w_{\rm eff} = 920$~Jy~ms, where $w_{\rm eff}^{\rm GP} =
8.8$~ms is the observed width of the GP. The energy of AP is $E =
S_{1752} \times P = 4.5$~Jy~ms. Then, the observed energy of the
GP exceeds the energy of the AP by factor of 200. This is the most
pronounced energy increase factor $k_{\rm E} =E^{\rm GP}/E^{\rm
AP}$ among the known pulsars with GPs. A pulse whose energy
exceeds the energy of the average pulse (AP) by more than a factor
of 100 is encountered approximately once in 3000 observed periods.

Lewandowski et al. (\cite{lewandowski}) show the bursting
behaviour of this pulsar, which spend 70-80\% of the time in a
"quasi-null" state. The amplitude of the AP formed only of
"burster" pulses will be about 4 times stronger that the AP
obtained by folding of all observed pulses. In this case the ratio
of the peak flux densities $S_{\rm peak}^{\rm GP}/S_{\rm
peak}^{\rm AP}$ and energies $E^{\rm GP}/E^{\rm AP}$ of the GP to
the AP will be about 4 times less. Even so, the observed energy of
the GP exceeds the energy of the AP by factor of 50, that is
nearly the same as for classical GPs in the Crab pulsar (see Table
1).

Along with a large intensity, the distinguishing characteristic of
the previously known pulsars with GPs is their two-mode pulse
intensity distribution.  At low intensities, the pulse strength
distribution is Gaussian, but above a certain threshold the pulse
strength distribution is roughly a power-law.

\begin{figure}
      \centering
      \resizebox{\hsize}{!}{\includegraphics{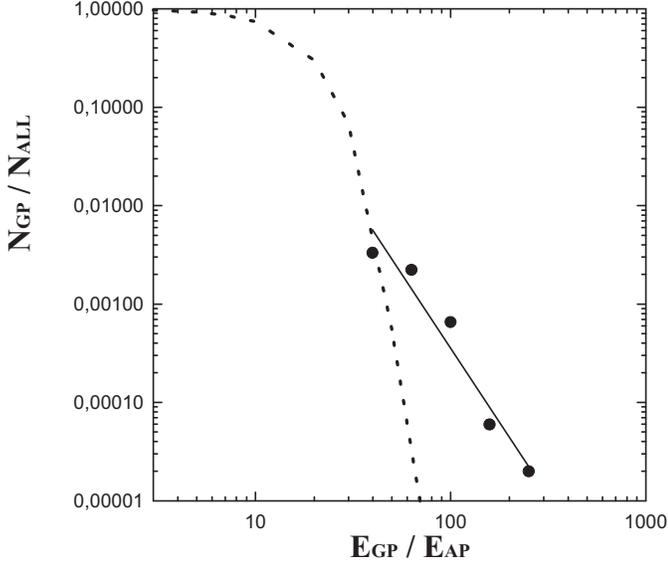}}
      \caption{The cumulative distribution of the observed
GP energy $E^{\rm GP}$ as related to the AP energy $E^{\rm AP}$.
The solid line is the observed power-law distribution $N^{\rm
GP}~/~N^{\rm All} \propto (E^{\rm GP}/E^{\rm AP})^{\alpha}$ with
index $\alpha = -3.0~\pm~0.4$. The dotted line represents the
possible version of the Gaussian distribution $N~/~N^{\rm All} =
exp(-a(E~/~E^{\rm AP})^2)$.
              }
         \label{Fig3}
   \end{figure}

In Fig.~\ref{Fig3} we show the measured cumulative distribution of
the ratio of the observed GP energy to the AP energy $E^{\rm
GP}_{\rm peak}/E^{\rm AP}_{\rm peak}$ of all pulses analyzed. In
the interval of $E^{\rm GP} / E^{\rm AP}$ from 40 to 200, the
distribution is fit by a power-law dependence $N^{\rm GP}/N^{\rm
All} \propto (E^{\rm GP}/E^{\rm AP})^{\alpha}$ with index $\alpha
= -3.0~\pm~0.4$.
%
%
For $E^{\rm GP}/E^{\rm AP}$ less than 40 the signal-to-noise ratio
is less than 5, and the observed distribution is masked by noise.
For this region  we plot  one of the possible versions of the
Gaussian distribution $N~/~N^{\rm All} = exp(-a(E~/~E^{\rm
AP})^2)$, which is tangent to the observed power-law distribution.

The intrinsic fine structure of GPs is masked by dispersion pulse
broadening $\Delta t_{\rm DM} = 4.4$~ms and a receiver time
constant $\tau = 3$~ms. Therefore, from October 5 through October
25, 2004, we performed additional observations with higher
temporal resolution. We used a 128-channel receiver with a channel
bandwidth of 1.25 kHz, sampling interval 0.81, and time constant
$\tau = 1$~ms. In this mode, we performed 14 observation sessions
containing 6800 pulsar periods.

\begin{figure}
      \centering
      \resizebox{\hsize}{!}{\includegraphics{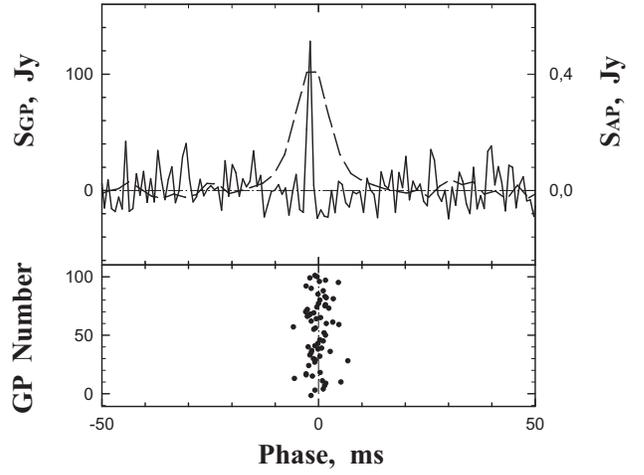}}
      \caption{ \textbf{(Top)} The  observed high resolution GP
      (bold line) and the AP (dotted line). The observed peak flux density
      of this GP exceeds the peak flux density of the AP by a factor of 320.
      The plot of the AP is presented on a 250 times larger
      scale and flux densities of the observed GP and AP
      are shown separately on the left and right sides of the "y"-axis.
       \textbf{(Bottom)}The phases of the GPs.
              }
         \label{Fig4}
   \end{figure}

Figure~\ref{Fig4}a shows the  high resolution GP (bold line)
together with the AP (dotted line). The observed peak flux density
of this GP exceeds the peak flux density of the AP by a factor of
320. The plot of the AP is presented on a 250 times larger scale,
and flux densities of the observed GP and AP are shown separately,
on the left and right sides of the "y"-axis.

The observed width of the GPs is $w_{0.5}^{\rm GP} = 1$ ms, which
is narrower than the AP by a factor of about 10.

Figure~\ref{Fig4}b shows the phases of the GPs whose intensities
exceeded the AP by a factor of 100. GPs cluster in a narrow phase
window near the middle of the AP. The clustering is closer for
stronger GPs.
%
%

The brightness temperature of the GPs is
$$
  T_{\rm B} = S \lambda^2 /2k\Omega~,
$$
where $\lambda$ is the radio wavelength, $k$ is the Boltzmann
constant, and $\Omega$ is the solid angle of the radio emission
region. Adopting $\Omega \simeq (l/d)^2$, where $l$ is the size of
the radio emission region  and $l \leq  c \times w_{0.5}$, where
$c$ is the speed of light and the distance to the pulsar $d = 2.7$
kpc (Lewandowski et al. \cite{lewandowski}), we obtain $T_{\rm B}
\geq 2 \times 10^{28}$ K.

\section{Discussion}

\begin{table*}[t]
\caption{The energy properties of giant pulses} \label{table:1}
\centering
\begin{tabular}{l c r c c c c c c l}     
\hline \hline 
PSR & $B_{LC}$ & Period & Freq & $E^{AP}$ & $S_{peak}^{GP}$ &
$W^{GP}$ &
$E^{GP}$ & $E^{GP}/E^{AP}$ & References \\
  & [G]       & [sec] & [MHz]&[Jy ms]   & [Jy]            & [ms]     & [Jy ms]  &           &     \\
\hline
J1752+2359 & 71 &  0.409 &  111 & 4.5   & 105    & 8.8     & 920  & 200     & this paper  \\
\hline
B0031--07 & 7.0 & 0.943 & 111 & 330 & 530            & 5  & $2.5\cdot10^3$  & 8  & Kuzmin et al. 2004  \\
          &     &       &  40 & 460 & $1.1\cdot10^3$ & 6  & $6.6\cdot10^3$  & 14  & Kuzmin \& Ershov 2004\\
\hline
J0218+4232  &$3.2\cdot10^{5}$ &$2.32\cdot10^{-3}$& 610& -  & -  & - & 1.34 & 51  & Joshi et al. 2004 \\
\hline
B0531+21 &$9.8\cdot10^{5}$ &$3.31\cdot10^{-2}$ &  594 & 6.4   & $150\cdot10^3$ & $5\cdot10^{-4}$  & 75    & 12    & Kostyuk et al. 2003 \\
          &     &   & 2228 & 0.11  & $18\cdot10^3$  & $5\cdot10^{-4}$  & 9     & 80   &        Kostyuk et al. 2003 \\
          &     &   & 5500 & 0.007 & $2.5\cdot10^3$   & $2\cdot10^{-6}$  & .005  & 0.7 &        Hankins et al. 2003\\
\hline
B1112+50  & 4.2 &  1.656 &  111 & 90    & 180    & 5     & 1440  & 16     & Ershov \& Kuzmin 2003\\
\hline
J1824--24 &$7.4\cdot10^{5}$ &$3.05\cdot10^{-3}$ & 1517 & $9.3\cdot10^{-3}$& -    & -   & 0.76  & 81    & Romani \& Johnston 2001\\
\hline
J1823-3021 &$2.5\cdot10^5$ &$5.44\cdot10^{-3}$ &  685 & -     & 45      &0.020   & -     &  64  & Knight et al. 2005\\
           &               &                   & 1400 & -     & 20      &0.007   & -     &  28  & Knight et al. 2005\\
\hline
B1937+21 &$1.0\cdot10^{6}$  &$1.56\cdot10^{-3}$ &  111 & 6.0   & -      & -     & 350   & 60   & Kuzmin \& Losovsky 2002\\
          &                 &                   &  430 & 0.4   & -      & -     & 7.5   & 19   & Kinkhabwala \& Thorset 2000\\
          &                 &                   & 1400 & 0.025 & -      & -     &  1    & 40   & Kinkhabwala \& Thorset 2000\\
\hline
B1957+20 & $3.8\cdot10^{5}$  &$1.61\cdot10^{-3}$ &  610 & -    & -      & -     & 0.925 & 129  & Joshi et al. 2004\\
\hline
\end{tabular}
\end{table*}

The GPs that we detected in PSR J1752+2359  exhibit all
characteristic features of the classical GPs in PSR B0531+21 and
PSR B1937+21.

The peak intensities of the GPs exceed the peak intensity of the
AP by more than a factor of 50. The histograms of the flux density
have a power-law distribution. The GPs are much narrower than the
AP and their phases are stable inside the integrated profile.

The most important aspect of this paper is the fact that PSR
J1752+2359 (as well as PSR B0031--07 and PSR B1112+50) represent
pulsars with a relatively low magnetic field $B_{\rm LC}$ at the
light cylinder. This is in contrast to the canonical suggestion
that GPs occur in pulsars with strong magnetic fields at the light
cylinder (e.g. Romani \& Johnston \cite{romani}). The detection
and first searches for GPs were performed in pulsars with
extremely high magnetic fields at the light cylinder of $B_{\rm
LC} = 10^4 - 10^5$~G. Then it was suggested that GPs originate
near the light cylinder (Istomin \cite{istomin}).  However,
detection of GPs in the pulsars PSR B1112+50 (Ershov \& Kuzmin
2003), PSR B0031--07 (Kuzmin, Ershov \& Losovsky 2004) and
presently reported GPs in PSR J1752+2359 have revealed that GPs
exist also in pulsars with ordinary magnetic fields at the light
cylinder of $B_{\rm LC} = 1 - 100$~G. These GPs may be associated
with the inner gap emission region (Gil \& Melikidze \cite{gil},
Petrova \cite{petrova}).

In Table 1 we summarize the comparative data on the energy
properties of giant pulses, for which the data on $E_{\rm GP}$ or
$E^{\rm GP}/E^{\rm AP}$ has been published or may be derived. All
known GPs, both with high and low magnetic field at the light
cylinder, have the same order of magnitude energy increasing
factor $k_{\rm E} =E^{\rm GP}/E^{\rm AP}$. Moreover, the presently
reported GPs of PSR J1752+2359 with relatively low magnetic field
at the light cylinder have the most pronounced $k_{\rm E} \cong
200$ among known pulsars with GPs.

An interesting aspect is the fine structure of GP radio emission.
Giant pulses from the Crab pulsar PSR B0531+21 (Sallmen et al.
\cite {sallmen}; Jessner et al. \cite {jessner}) and LMC pulsar
PSR B0540-69 (Johnston \& Romany, \cite {johnston}) frequently
show a double structure of GPs. Kuzmin \& Ershov (\cite {kuzmin})
have detected a double structure of GPs from the pulsar PSR
B0031-07 and argued that the pulses come from the polar region and
not near the outer magnetosphere. In contrast Kinkhabwala \&
Thorsett (\cite {kinkhabwala}) reported that after inspecting many
pulses from PSR B1937+21 they see no evidence for multiple-peaked
emission. Soglasnov et al. (\cite{soglasnov}) also claimed that
GPs from this pulsar occur, in general, with a single spike.

In our observations of PSR J1752+2359 we used a 128-channel
receiver with a channel bandwidth of 20 kHz. In this mode the
intrinsic fine structure of GPs was masked by the dispersion
broadening $\Delta t_{\rm DM} = 4.4$~ms and receiver time constant
$\tau = 3$~ms which makes it hard to look for a double structure
of GPs.

We have attempted to look for a fine structure of GPs in the
additional observations with higher temporal resolution using
128-channel receiver with a channel bandwidth of 1.25 kHz,
sampling interval 0.81 ms and time constant $\tau = 1$~ms. But in
this mode the radio telescope sensitivity was worse by factor of 4
and we performed only 14 observation sessions containing 6800
pulsar periods against 120 observation sessions containing 50400
pulsar periods in the main mode. Therefore in the high temporal
resolution mode we detected only 5 GPs, but no double structure.

One should note that the GPs of PSR J1752+2359 with relatively low
magnetic field at the light cylinder was observed at the low
frequency of 111 MHz, whereas GPs of pulsars with strong magnetic
fields at the light cylinder were observed mainly at high
frequencies. It would be of interest to observe  GPs of PSR
J1752+2359 at high frequencies, where one can obtain better
temporal resolution.

\section{Conclusions}

Giant Pulses (GPs) from pulsar PSR J1752+2359 have been detected.
The energy of the GPs  exceeds the energy of the average profile
by a factor of up to 200, which is greater than in other known
pulsars with GPs. The cumulative distribution is fit by a
power-law with index $-3.0~\pm~0.4$. PSR J1752+2359 as well as the
previously detected PSR B0031--07 and PSR B1112+50 are the first
pulsars with GPs that do not have a high magnetic field at the
light cylinder.

 \begin{acknowledgements}

We wish to thank  V.~V. Ivanova, K.~A. Lapaev \& A.~S. Aleksandrov
for assistance during observations.  This work was supported in
part by the Russian Foundation for Basic Research (project No
05-02-16415) and the Program of the Presidium of the Russian
Academy of Sciences "Non-steady-state Processes in Astronomy". We
are grateful to Dr. Steve Shore for valuable comments and
suggestions.

\end{acknowledgements}


\end{document}